\documentclass[sigconf]{acmart}

\usepackage{booktabs} 
\usepackage{algorithm}
\usepackage{algorithmic}
\usepackage{graphicx}
\usepackage{subfigure}

\copyrightyear{2019}
\acmYear{2019} 
\setcopyright{iw3c2w3}
\acmConference[WWW '19]{Proceedings of the 2019 World Wide Web Conference}{May 13--17, 2019}{San Francisco, CA, USA}
\acmBooktitle{Proceedings of the 2019 World Wide Web Conference (WWW'19), May 13--17, 2019, San Francisco, CA, USA}
\acmPrice{}
\acmDOI{10.1145/3308558.3308558.3313622}
\acmISBN{978-1-4503-6674-8/19/05}

\usepackage{balance}

\begin{document}

\title{Tag2Vec: Learning Tag Representations in Tag Networks}

\author{Junshan Wang}
\affiliation{%
  \institution{Key Laboratory of Machine Perception, Ministry of Education, Peking University}
  \city{Beijing}
  \state{China}
  \postcode{100871}
}
\email{wangjunshan@pku.edu.cn}

\author{Zhicong Lu}
\affiliation{%
  \institution{Key Laboratory of Machine Perception, Ministry of Education, Peking University}
  \city{Beijing}
  \state{China}
  \postcode{100871}
 }
\email{phyluzhicong@pku.edu.cn}

\author{Guojie Song}
\authornote{Corresponding author.}
\affiliation{%
  \institution{Key Laboratory of Machine Perception, Ministry of Education, Peking University}
  \city{Beijing}
  \state{China}
  \postcode{100871}
 }
\email{gjsong@pku.edu.cn}

\author{Yue Fan}
\affiliation{%
  \institution{Key Laboratory of Machine Perception, Ministry of Education, Peking University}
  \city{Beijing}
  \state{China}
  \postcode{100871}
 }
\email{fanyue@pku.edu.cn}

\author{Lun Du}
\affiliation{%
  \institution{Key Laboratory of Machine Perception, Ministry of Education, Peking University}
  \city{Beijing}
  \state{China}
  \postcode{100871}
 }
\email{dulun@pku.edu.cn}

\author{Wei Lin}
\affiliation{%
  \institution{Alibaba}
  \city{Beijing}
  \state{China}
}
\email{yangkun.lw@alibaba-inc.com}

\begin{abstract}
Network embedding is a method to learn low-dimensional representation vectors for nodes in complex networks. In real networks, nodes may have multiple tags but existing methods ignore the abundant semantic and hierarchical information of tags. This information is useful to many network applications and usually very stable. In this paper, we propose a tag representation learning model, Tag2Vec, which mixes nodes and tags into a hybrid network. Firstly, for tag networks, we define semantic distance as the proximity between tags and design a novel strategy, parameterized random walk, to generate context with semantic and hierarchical information of tags adaptively. Then, we propose hyperbolic Skip-gram model to express the complex hierarchical structure better with lower output dimensions. We evaluate our model on the NBER U.S. patent dataset and WordNet dataset. The results show that our model can learn tag representations with rich semantic information and it outperforms other baselines. 
\end{abstract}

%
%



\maketitle

\section{Introduction}

Network embedding is a method to learn low-dimensional representation vectors for nodes in complex network and has drawn much attention recently \cite{perozzi2014deepwalk, grover2016node2vec, tang2015line}. The representation vectors helps to complete many machine learning tasks, such as node classification \cite{bhagat2011node}, node clustering \cite{mishra2007clustering} and link prediction \cite{lu2011link}.

But in real networks, nodes are often attached with multiple tags, and each tag represents a characteristic of the node. For example, there are interested domain tags of researchers in academic networks and discussion topic tags of users in forum networks \cite{leskovec2005graphs}. Tags have abundant semantic information, which can be used to measure the semantic distances between tags. For example, in academic networks, considering the semantic distance between two interested domain tags, Artificial Intelligence is closer to Machine Learning than Operating System. In addition, tags have abundant hierarchical information. For example, Optics and Transportation are the sub-categories of Mechanical, which should be considered when leaning their representations. But considering tags as discrete features with one-hot encoding in previous work is based on a simple assumption that different features are irrelevant. It will also cause the sparsity of vectors and inefficiency of algorithms.

Therefore, similar to learning the node representation vectors, we want to learn the tag representation vectors in networks, so that the semantic distances between tags can be reflected. 
On the one hand, because of the semantic and hierarchical information of tag representations, they can help to improve the performance of network applications and study the characteristics and behaviors of tags. On the other hand, as tags are usually stable, we can obtain good representations even though part of networks is lacking. 

Existing network embedding methods, such as DeepWalk \cite{perozzi2014deepwalk} and LINE \cite{tang2015line}, only learn the node representations. But tag networks are a special kind of heterogeneous networks which contains the interaction relationship between nodes as well as the affiliation relationship of tags and nodes. There are two main challenges of learning tag representations in the heterogeneous networks:
\begin{itemize}
    \item \textit{How to learn the semantic information of tags?} If two nodes have tags with similar semantic information, they will have small semantic distance. If two tags have many members with close relationship, they will be close as for semantic distance. So in tag representation learning, we have to consider the semantic information of nodes and tags. Besides, we need to distinguish the interaction relationship between nodes and the affiliation relationship between tags.
    \item \textit{How to represent the tags?} In tag representation, we describe the complex traverse and vertical structure of tags with the help of interaction relationships between nodes and affiliation relationships between nodes and tags. Traditional embedding methods learn interaction relationships in Euclidean spaces, but ignore affiliation relationships. If we use tree-like structure to describe the data in Euclidean spaces, it will causes the problem of high computational cost and data sparsity when the hierarchical structure becomes complex. 
\end{itemize}

In this paper, we propose a tag representation learning model, Tag2Vec. We learn the representation vectors of tags to reflect their semantic information and hierarchical structure. Firstly, we mix tags and nodes into a hybrid network to consider the influence of semantic information between tags and nodes. We adopt the parameterized random walk strategy to generate the context that reflects the semantic distance between tags. Then, because of the disability of learning vertical affiliation relationship in Euclidean spaces, we propose the hyperbolic Skip-gram model to learn representation vectors in hyperbolic spaces. Hyperbolic spaces are a kind of metric spaces, and they expand exponentially with regard to the radius $r$. They confirm to the growth of hierarchical relationship of tags so that learning representations in hyperbolic spaces can better describe the affiliation relationship of tag networks. 

The main contribution of this paper is:
\begin{itemize}
    \item We propose a tag representation learning model Tag2Vec, which learns tag representation vectors in networks to reflect the semantic information of tags. 
    \item We design a parameterized random walk strategy to generate the context with semantic information. It sets parameters to control the proportion of different semantic information. 
    \item We propose hyperbolic Skip-gram model to learn representations in hyperbolic spaces. We improve the representative ability of tag vectors with the help of two advantages of hyperbolic spaces: capable to learn affiliation relationship and reduce the dimensions. 
\end{itemize}
We conduct experiments on two datasets. The results demonstrate that our model can learn representations with rich semantic information of tags and is superior than baselines.

\section{Related Work}
As tags are usually text or labels, they can be regarded as special nodes in heterogeneous networks or attributes of nodes. Meanwhile, the hierarchical structure of tags can be expressed in hyperbolic spaces. So we discuss related work of three kinds of network embedding methods \cite{cui2018survey} separately: structure-preserving, attribute-preserving and hyperbolic network embedding.

Many successful network embedding methods have been proposed recently for preserving topological structure \cite{perozzi2014deepwalk, grover2016node2vec, tang2015line, cao2015grarep, wang2016structural, du2018dynamic}. They are working on a homogeneous network but insufficient in networks with tags for the ignorance of tag information. 
A feasible method is to build heterogeneous networks, where tags are regarded as another type of nodes. Existing heterogeneous network embedding methods can be achieved by partitioning heterogeneous networks into a set of networks \cite{tang2015pte, xu2017embedding}, establishing a deep learning structure \cite{chang2015heterogeneous, wang2018shine}, or using meta path based models \cite{dong2017metapath2vec, fu2017hin2vec, huang2017heterogeneous}. However, these methods ignore the higher status of tags in hybrid networks and also leave the hierarchy between tags hidden. Therefore, Tag2Vec is proposed in order to represent tags with both semantic and hierarchical information clearly. 

Attribute-preserving network embedding methods embed nodes according to both the graph topology and the node attributes. An attribute of a node can be a text \cite{yang2015network}, a label \cite{huang2017label,pan2016tri}, or the node content \cite{sun2016general,pan2016tri}. If two nodes have similar attributes, they will have close embedding vectors. 
However, attribute-preserving network embedding does not give every attribute an embedding vector and does not learn semantic information between attributes explicitly. Tag2Vec learns the representations of tags and nodes simultaneously, modelling the relationship between nodes and between tags.

Many real world networks actually have hierarchical or tree-like structure \cite{adcock2013tree}. Learning representations in Euclidean spaces requires large dimensions to hold a network with complex hierarchical relationships\cite{nickel2014reducing, bouchard2015approximate}.
Hyperbolic space \cite{gromov1987hyperbolic,boguna2010sustaining} can reflect the latent hierarchy of embedded nodes with low dimensions. Poincare network embedding \cite{nickel2017poincare} embeds words into hyperbolic space in a way such that their distance in the embedding space reflects their semantic similarity. \cite{pmlr-v80-ganea18a} embeds acyclic graphs into hyperbolic space to learn their hierarchical features. But these methods only focus on learning representations in homogeneous networks without tags. In this work, we learn the representations of both nodes and tags under the hyperbolic space.

\section{Problem Definition}
In this section, we define some problems in tag embedding formally.

\begin{definition}
A \textbf{Tagged Network} is defined as $G=<V,E,T>$, where $V=\{v_i\}, i=1,2,...,n$ represents the set of nodes, $E=\{e_{ij}\}, i,j=1,2,...,n,i \neq j$ represents the set of edges, and $T=\{T_i\}, i=1,2,...,n$ represents the set of tags that each node belongs to. Each tag corresponds to a literal symbol and labels a characteristic of a a node. Each node $v_i$ have $k$ tags $T_i = \{t_{i_1}, t_{i_2}, ..., t_{i_k}\}, t_{i_j} \in T$. If node $v_i$ is labeled by tag $t$, node $v_i$ has a tag $t$.
\end{definition}

In a tagged Network, an edge between two nodes reflects their interaction relationship, while an edge between a node and a tag reflects their affiliation relationship. For example, in social networks, a node is a user, and two users have friend relationship if there exists an edge between them. A tag of the node may be a characteristic label, an interested topic or a community of the user. A node corresponds to multiple tags and a tag corresponds a set of nodes. 

The text of a tag can only reflect part of the semantic information. Its semantic information is decided by the characteristic of its nodes in spirit. Therefore, we introduce ground-truth community to define the semantic distance between tags.

\begin{definition}
A \textbf{Ground-truth Community} $ C_t$ is the set of nodes labeled by tag $t$.
\end{definition}

A tag corresponds to a ground-truth community. The distance between tags is defined by the relationship between their ground-truth communities. In the following, we introduce two types of relationships between Ground-truth Communities.

\begin{definition}
\textbf{Member Similarity} is the ratio of the common members to all members in two ground-truth communities, which is extended from the definition of common neighbor similarity \cite{du2018galaxy}:
$$ D_{MS}(C_{t_1},C_{t_2}) = \frac{|C_{t_1} \cap C_{t_2}|^2}{|C_{t_1}| \cdot |C_{t_2}|}. $$
\end{definition}

For two tags in a tagged network, they are closer semantically if their ground-truth communities have more common members. 

\begin{definition}
\textbf{Social Closeness} is the ratio of the connected edges between members of two ground-truth communities to all their inter-connected edges:
$$ D_{SC}(C_{t_1},C_{t_2}) = \frac{|Edge(C_{t_1},C_{t_2})|^2}{|Edge(C_{t_1},V-C_{t_1})| \cdot |Edge(C_{t_2},V-C_{t_2})|}, $$
where $Edge(C_{t_1},C_{t_2})$ is the set of the connected edges between members of $C_{t_1}$ and $C_{t_2}$, and $Edge(C_{t_1},V-C_{t_1})$ is the set of all inter-connected edges of $C_{t_1}$.
\end{definition}

For two tags in a tagged network, they may be close semantically if the members of their ground-truth communities have close relationship.

\begin{definition}
Given a tagged network $G=<V,E,T>$, the problem of \textbf{Tag Embedding} aims to represent each tag $t \in T$ in a low-dimensional space $\mathbb{R}^d$. In other words, it aims to learn a mapping function $f: T \to \mathbb{R}^d$, where $d<<|V|$, and the representation vectors in $\mathbb{R}^d$ preserve both member similarity and social closeness between ground-truth communities.
\end{definition}

The computational complexity of member similarity and social closeness is $\Theta(|T|^2)$, where $|T|$ is the number of tags. As the complexity of social closeness is related to node numbers of communities, average degrees of communities, and neighborhood orders, it will be much higher in realit. Therefore, to avoid the expensive computation, we mix the nodes and tags into the same network to learn their semantic information, including the member similarity and social closeness between the tags. And we call the mixed network as node-tag hybrid network:

\begin{definition}
A \textbf{Node-Tag Hybrid Network} $H=<V',E'>$ is derived from a tagged network $G=<V,E,T>$, where $V'=V+T, E'=E+E_T$. For $v' \in V'$, $v'$ is a plain node if $v' \in V$. Otherwise, it is a tag node if $v' \in T$. For $e' \in E'$, $e'$ is an interaction edge if $e'=<v_1,v_2> \in E$. Otherwise, it is an affiliation edge if $e'=<v,t>$ and node $v$ is labeled by tag $t$.
\end{definition}
\section{Tag Representation Learning Model}
If we regard a node-tag hybrid network as a normal network and use traditional network embedding methods, we can get the representations for both plain nodes and tag node. However, the difference between plain nodes and tag nodes in the Node-Tag Hybrid Network are ignored. And we can't learn the semantic information of tags explicitly, neither.

In this section, we propose Tag2Vec, a tag representation learning model. In a node-tag hybrid network, we design a new strategy, parameterized Random walk, to generate the context with both member similarity and social closeness information between tags. Meanwhile, We train a hyperbolic Skip-gram model with the generated context to learn the hyperbolic representation vectors of plain nodes and tag nodes. It can reduce the output dimensions and describe the transverse and vertical relationship between tags more accurately.

\subsection{Parameterized Random Walk}
We design a Tag2Vec parameterized random walk strategy in a node-tag hybrid network. Firstly, it can reflect both member similarity and social closeness between tags, and the ratio between their weights can be controlled by the parameter. Secondly, it should contain hierarchical information of both transverse and vertical relationship between tags for hyperbolic Skip-gram model, and the ratio between their weights can be controlled by the parameter, too. Specifically, we adopt two strategies to control parameters for different semantic information and different dimensional information.

\subsubsection{Walk Patterns for Semantic Information}
If two tags have similar semantic information, their representation vectors should be closer in the vector space. So, tag nodes with smaller semantic distance should appear more simultaneously in the context. As the semantic distance is decided by member similarity and social closeness, we consider two basic walk patterns for them respectively.

\begin{figure}[!htbp]
\centering
\begin{minipage}[b]{0.5\textwidth}
\centering
\subfigure[MS-driven walk. ] { 
\label{fig:ms} 
\includegraphics[width=0.4\columnwidth]{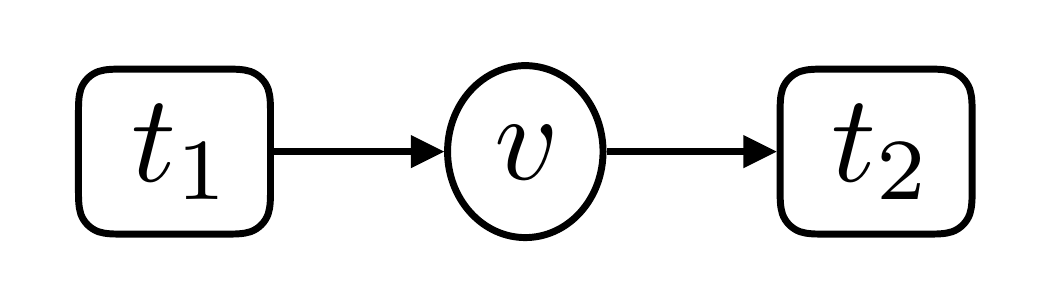}
} 
\subfigure[SC-driven walk.] { 
\label{fig:sc} 
\includegraphics[width=0.5\columnwidth]{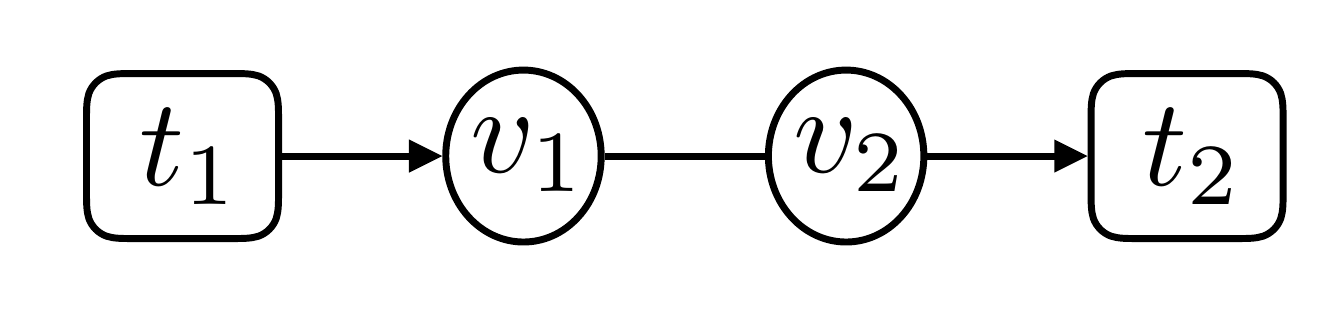}
} 
\caption{Walk Patterns.} 
\label{fig:walk_patterns} 
\end{minipage}
\end{figure}

\begin{itemize}
    \item \textbf{MS-driven walk.} As shown in Figure \ref{fig:ms}, in MS-driven walk, tag nodes with common members may appear simultaneously in the random walk sequences. There must be one and only one plain node between a pair of tag nodes, and the plain node is the common member of two tags. 
    \item \textbf{SC-driven walk.} As shown in Figure \ref{fig:sc}, in SC-driven walk, tag nodes with connected members may appear simultaneously in the random walk sequences. There must be more than two plain nodes between a pair of tag nodes, and the adjacent plain nodes are connected.
\end{itemize}

We notice that in a node-tag hybrid network, there are no edges between tag nodes. Combining the two basic walk patterns, we find that when the current node is a tag node, the next node must be a plain node. When the current node is a plain node, the next node may be a tag node or a plain node. 

When the current node is a tag node, we should select a member node from the ground-truth community of the tag. We consider that members with high social influence are more important to the community. So we calculate the weights of tag-to-plain edges according to the centrality, such as degree centrality, of plain nodes in the original network.

When the current node is a plain node, we should select the type of nodes as the next step, a tag node or a plain node. A tag node corresponds to MS-driven walk, and a tag node corresponds to SC-driven walk. So we set a parameter $p$ as the probability to select a tag node and adopt the strategy of MS-driven walk.

\subsubsection{Transverse and Vertical Relationship}
As we discussed above, we introduce the hyperbolic representation vectors to distinguish the transverse interaction and vertical affiliation relationships, and discover the hierarchical semantic information of tags. This requires that the context generated by random walk must contain enough transverse and vertical relationships between tags. However, in a node-tag hybrid network, there are no edges between tags, and the correlation information between tags is obtained implicitly by the random walk sequence. We can know which type of relationship two tags have only when we finish the learning procedure. So we want to control the relationship between tags in the same sequence and the ratio between two types of relationships.

We find that the sizes of ground-truth communities are effective to infer the relationship between tags. If two tags have transverse interaction relationship, the difference between their community size may be small. While two tags with vertical affiliation relationship, the difference between their community size may be large. Thus, we can control the selection of the next tag in the random walk sequence according to their community sizes. 

Specifically, in the random walk, if the current node is a plain node, we have to select a tag node from the tag neighbors as the next tag $t$. We consider the last tag $t_{last}$ in the current random walk sequence and compare the community size of $t_{last}$ with $t$. We consider two strategies.
\begin{itemize}
    \item \textbf{Transverse Interaction Relationship.} If the next step is to sample for transverse social relationship, we tend to select the tags having similar community size with the last tag.
    $$ W_{simi}(v,t) = B \cdot \exp( -(|t| - |t_{last}|)^2 )$$
    \item \textbf{Vertical Affiliation Relationship.} If the next step is to sample for vertical affiliation relationship, we tend to select the tags having different community size with the last tag.
    $$ W_{diff}(v,t) = B \cdot \exp( (|t| - |t_{last}|)^2 )$$
\end{itemize}
Notice that $B$ is the normalization constant. We set a parameter $q$ as the probability to adopt the transverse interaction relationship.

\subsection{Hyperbolic Skip-Gram}
\begin{figure}[!htbp]
\centering
\begin{minipage}[b]{0.5\textwidth}
\centering
\subfigure[Tree Structure] { 
\label{fig:tree} 
\includegraphics[width=0.47\columnwidth]{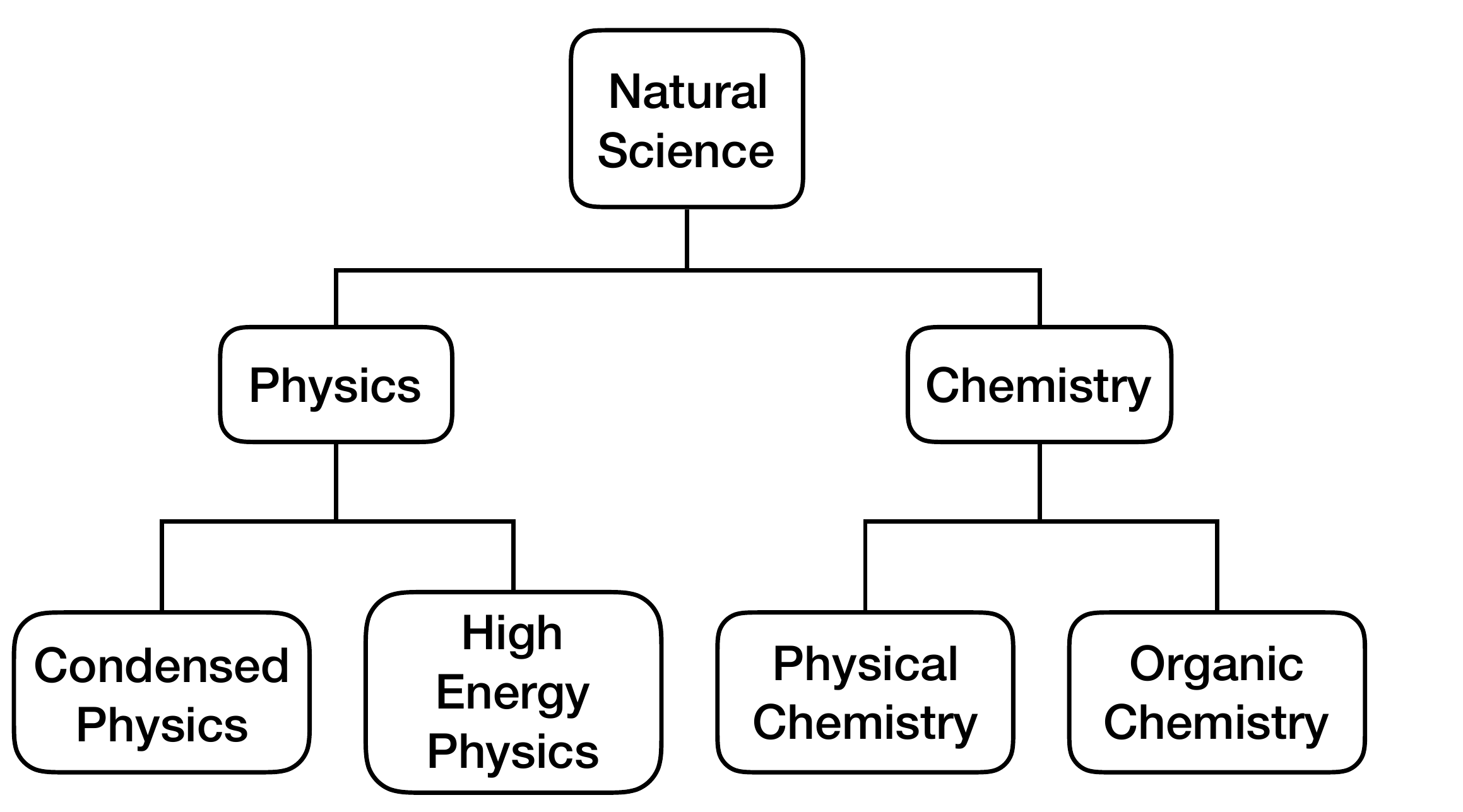}
} 
\subfigure[Hyperbolic Structure] { 
\label{fig:hyperbolic} 
\includegraphics[width=0.47\columnwidth]{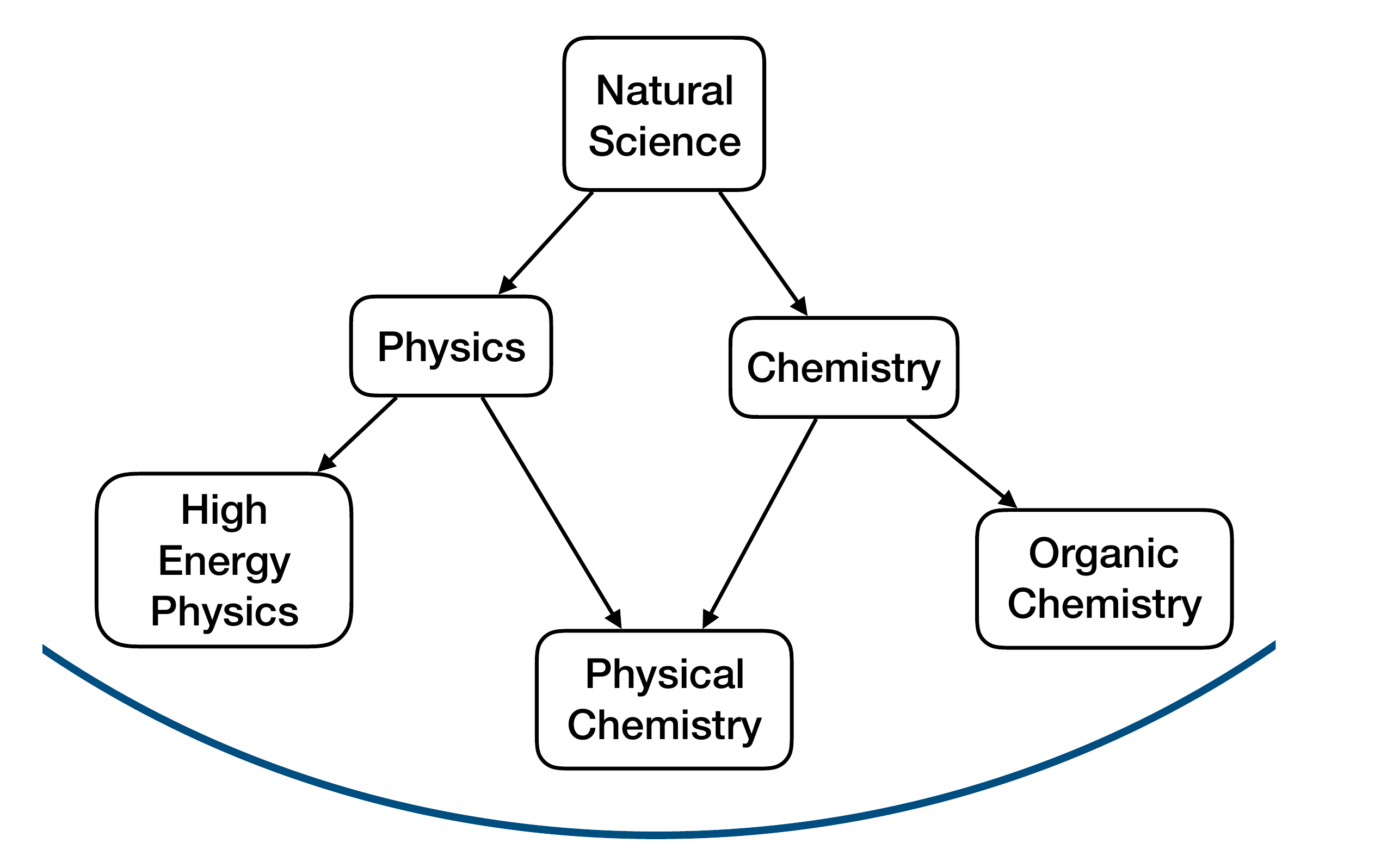}
} 
\caption{The Structure of A Tag Network} 
\label{fig:structure} 
\end{minipage}
\end{figure}

In tag representation learning, the relationships between tags contain both transverse interaction relationships and vertical affiliation relationships. But in fact, most tag relationship is between transverse and vertical. For example, in Figure \ref{fig:tree}, it is a tree describing the relationships between research fields. A node and one its children have affiliation relationship, while the children of a node may have interaction relationship. But as for two subjects, Physics and Physical Chemistry, we can't describe their relationship in this tree.

Hyperbolic geometry is a non-Euclidean geometry that studies the spaces of constant negative curvature. It is capable to model the hierarchical structure with low dimensions. Meanwhile, in continuous hyperbolic spaces, there are no distinct hierarchies, which means there are no distinct affiliation relationships. In general, nodes with higher levels are placed closer to the origin of the sphere and the children of a node are placed in the area with similar angle and longer distance to the origin. We show relationships discussed above in a two-dimensional hyperbolic space in Figure \ref{fig:hyperbolic}. It describes the relationships between Physical Chemistry, Physics and Chemistry suitably.
Besides, Euclidean spaces expand polynomially while hyperbolic spaces expand exponentially with regard to the radius $r$. So we can learn the representations in hyperbolic spaces with less parameters and lower dimensions.
Therefore, embedding in hyperbolic spaces can not only describe the complex relationship between tags, but also reduce the output dimensions. 

Existing network embedding method\cite{nickel2017poincare} uses a pure strategy that models the edge probability and it is not a Skip-gram procedure. That is to say that we can't learn the hyperbolic representation vectors from the context generated by Tag2Vec paratemeterized random walk. Thus, we propose a method to learn the representations on the Poincare ball with Skip-gram, as Poincare disk is a common and simple way to model htperbolic geometry.

Euclidean and hyperbolic spaces are both Riemannian spaces. Their difference is that Euclidean spaces have isotropic Riemannian metric tensor $g^{(E)}$. While in hyperbolic spaces, the metric tensor is:
\begin{equation}
\label{equ:metric_tensor}
    g^{(H)}_x = (\frac{2}{1-\|\vec x\|^2})^2 g^{(E)}.
\end{equation}

Therefore, in hyperbolic spaces, $g^{(H)}_x$ is related to the vector $\vec x$ and it is not shift-invariant. So there isn't a simple formation of global inner-product. But there is still the definition of the distance between two points in hyperbolic spaces:
\begin{equation}
\label{equ:distantce_definition}
d_H(\vec u, \vec v) = \cosh^{-1}(1 + \frac{\|\vec u - \vec v\|^2}{(1-\|\vec u\|^2)(1-\|\vec v\|^2)}).
\end{equation}

In traditional Skip-gram model, we use the scoring function $s(w,h)$ between the word $w$ and the context $h$ to evaluate their closeness. Then we have the probability distribution of the context $h$ given the word $w$ in the neural probabilistic language model \cite{bengio2003neural}
$
P(h|w) = \frac{\exp(s(w,h))}{\sum_{h'}\exp(s(w,h'))} .
$

Traditionally, $s(w,h)$ is a linear function, where we calculate the inner-product of $\vec{w}$ and $\vec{h}$. It is an operation in Euclidean spaces and we want to transfer it into hyperbolic spaces. But as mentioned above, there is no definition of inner-product in hyperbolic spaces. So we transform $s(w,h)$ into a function with regard to the distance:
\begin{equation} 
s(w,h) = -A\|\vec{w}-\vec{h}\|^2 + B = -A(d^2(\vec{w},\vec{h})) + B, 
\end{equation}
where $A$ and $B$ are parameters. But as the probability is normalized by a constant $\sum_{h'}\exp(s(w,h'))$, they have no influence on the model. So the probability distribution $P(h|w)$ can be calculated according to the Euclidean distance:
\begin{equation} P(h|w) = \frac{\exp(\Phi(d(w,h)))}{\sum_{h'}\exp(\Phi(d(w,h')))}, 
\end{equation}
where $\Phi$ can be a non-linear operator.

Therefore, we replace Euclidean distance $d_E$ with hyperbolic distance $d_H$. For simplicity, we let $\Phi$ as an identify function and define the loss function using maximum likelihood estimation. We use negative sampling to accelerate and the final loss function is
\begin{equation}
   \begin{aligned}
   L(\vec w) = - \sum_{(i, j) \in D} & \log \sigma(-d_H(\vec w_i , \vec h_j)) \\
   & + k \cdot \mathbb{E}_{h_n \sim P_n(h)} \left[ \log \sigma(d_H(\vec w_i, \vec h_n)) \right],
\end{aligned} 
\end{equation}
where $\sigma(x)=\frac{1}{1+\exp(-x)}$ and $P_{n}(h)$ is the noise distribution for negative sampling.

\subsubsection{Optimization}

As we use distance function $d_H(w,h)$ in hyperbolic spaces, the loss function can be optimized with Riemannian stochastic gradient descent (RSGD) optimization method \cite{bonnabel2013stochastic}.

\section{Experiment}
In order to demonstrate the superiority of our tag embeddings, we conduct the experiment on two datasets. 

\subsection{Dataset}
We have to choose datasets with tags of nodes, such as communities, categories, and other discrete features. Meanwhile, these tags may have implicit hierarchical correlation. 
\begin{itemize}
    \item \textbf{NBER Patent Citation Data} \cite{hall2001nber} describes patents in U.S. over the last 30 years. Each patent is a node and each citation is an edge in our network. Each node has one class, one category and sub-category as its tags. There are 6 categories, 35 sub-categories, and 426 classes classified by USPTO. We extract a sub-network from the dataset with 401 nodes, 462 tags and more than 3000 edges. To be notified, the affiliation relationship between categories and sub-categories is not part of the input and we will learn it from our model.
    \item \textbf{WordNet Data} \cite{miller1995wordnet} is a lexical database. It groups English words into sets of synonyms called synsets and records relations among these synonym sets or their members. We extract a subset from the dataset and construct a network. Each node in the network is a leaf in the tree and its tags are k-order ancestors and each edge in the network represents the genetic relationship between two words.
\end{itemize}

\subsection{Baselines}
We use the following baselines comparing with our model. Except the first baseline, we learn the representation vectors of nodes and tags simultaneously in a node-tag hybrid network. 
\begin{itemize}
    \item Origin: We construct a plain network without tags and learn the representation vectors of nodes by DeepWalk. Then we encode the tags into one-hot vectors, and concatenate the node vectors and node vectors together.
    \item DeepWalk \cite{perozzi2014deepwalk}: It generates context of nodes by random walks and adopts Skip-gram model.
    \item LINE \cite{tang2015line}: It preserves both first-order and second-order proximities between nodes. 
    \item Node2Vec \cite{grover2016node2vec}: It is an improvement of DeepWalk, and it adds more flexibility in exploring neighbors.
\end{itemize}

\subsection{Tag Representation Visualization}
\begin{figure}[ht]
	\includegraphics[width=0.6\linewidth]{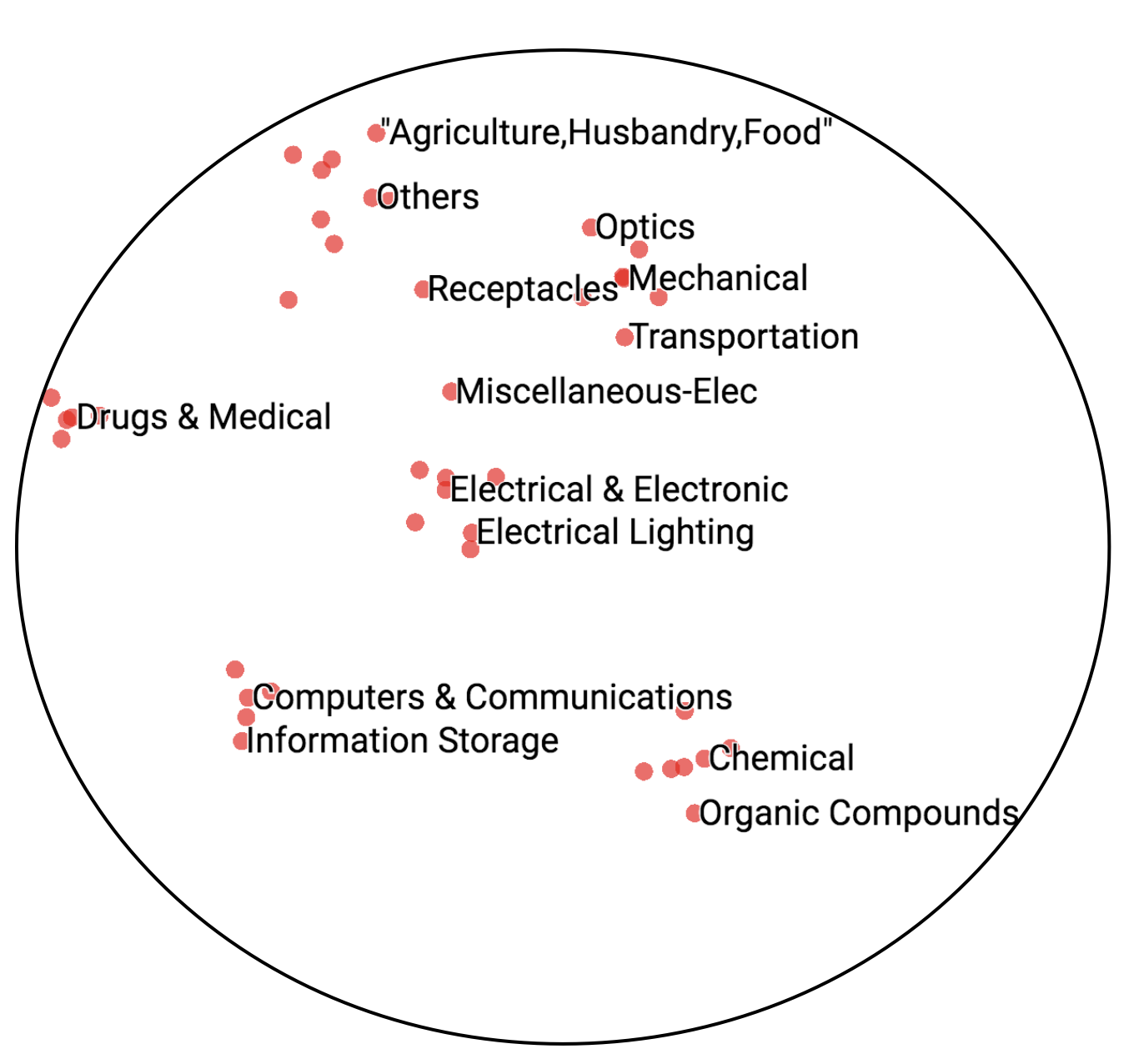}
	\caption{Tag Visualization} 
    \label{fig:tag_visualization} 
\end{figure}

Figure \ref{fig:tag_visualization} shows the visualization of tags learned from our model on NBER Patent Citation dataset. It is obvious that the tag representations learned from our model have good hierarchical property. 6 categories are distant from each other and dispersed in the space. Sub-categories of the same category are close to each other and cluster around the category. For example, Optics and Transportation are sub-categories of Mechanical, and they are close around Mechanical.

\begin{figure*}[ht]
	\begin{minipage}{0.35\linewidth}
		\includegraphics[width=1.0\linewidth]{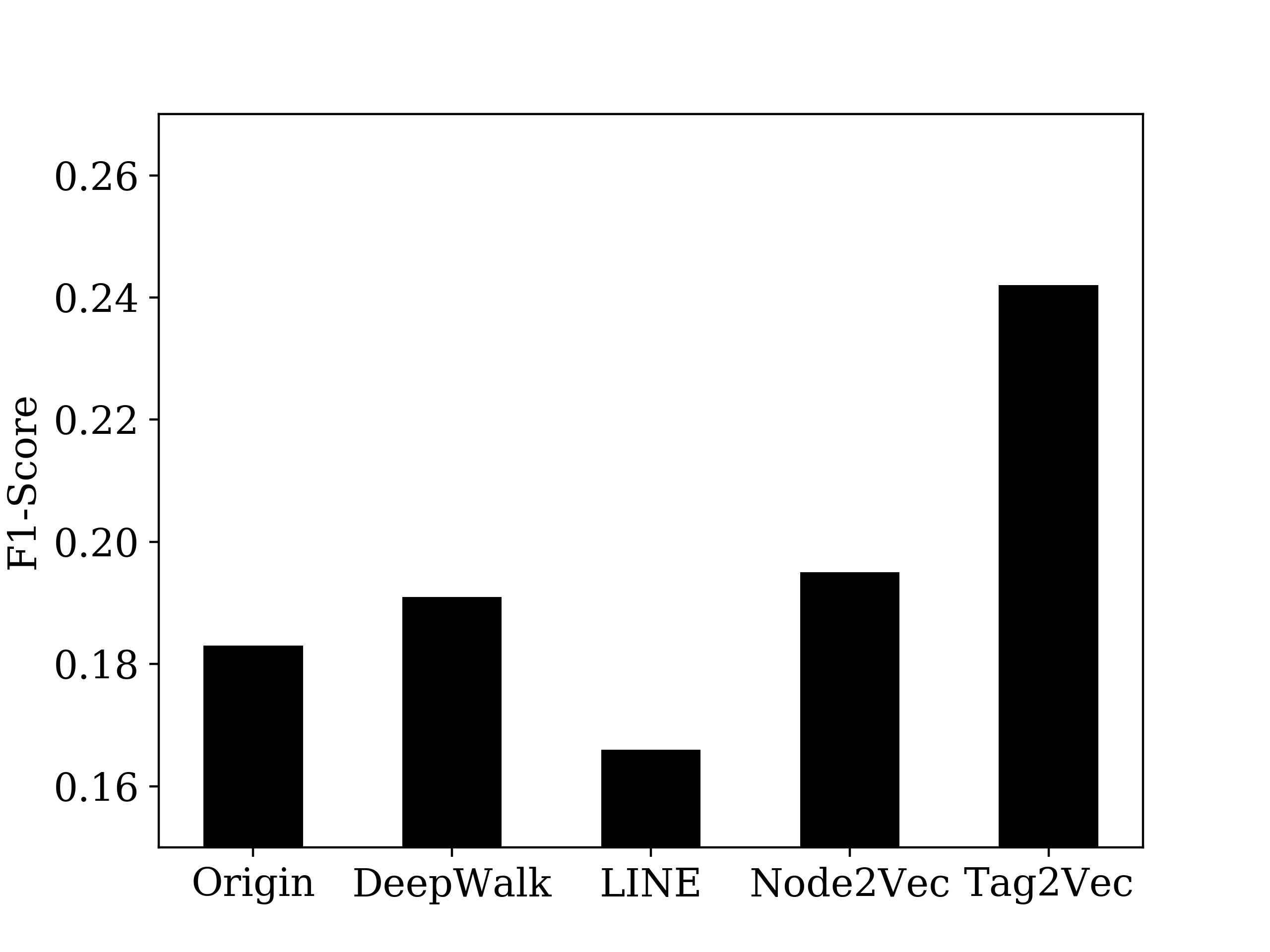}
		\caption{Node Classification} 
        \label{fig:node_classification} 
	\end{minipage}
	\begin{minipage}{0.35\linewidth}
		\includegraphics[width=1.0\linewidth]{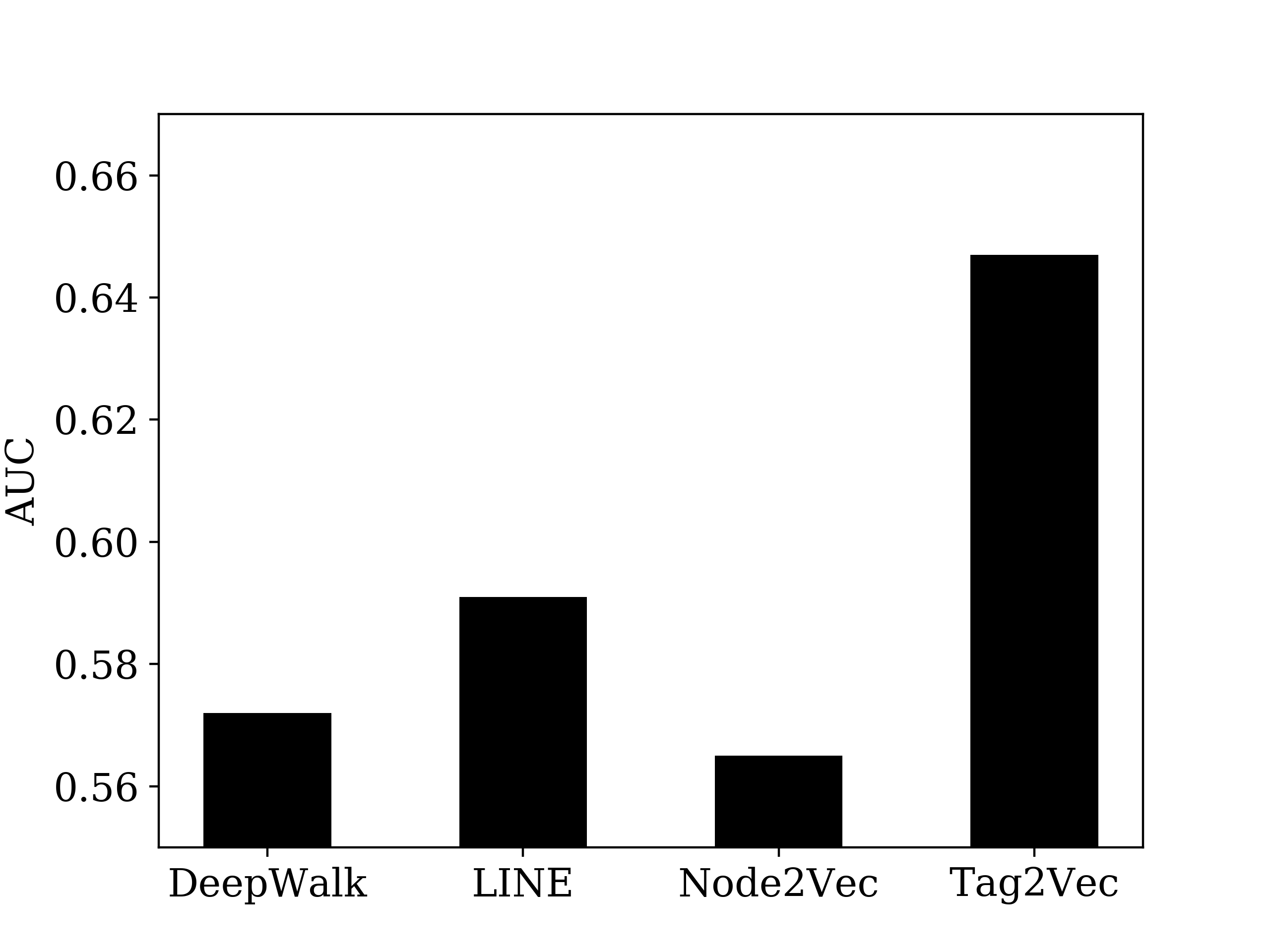}
		\caption{Similar Community Detection} 
        \label{fig:similar_community_detection} 
	\end{minipage}
\end{figure*}

\begin{figure*}[ht]
    \begin{minipage}{0.33\linewidth}
		\includegraphics[width=1.0\linewidth]{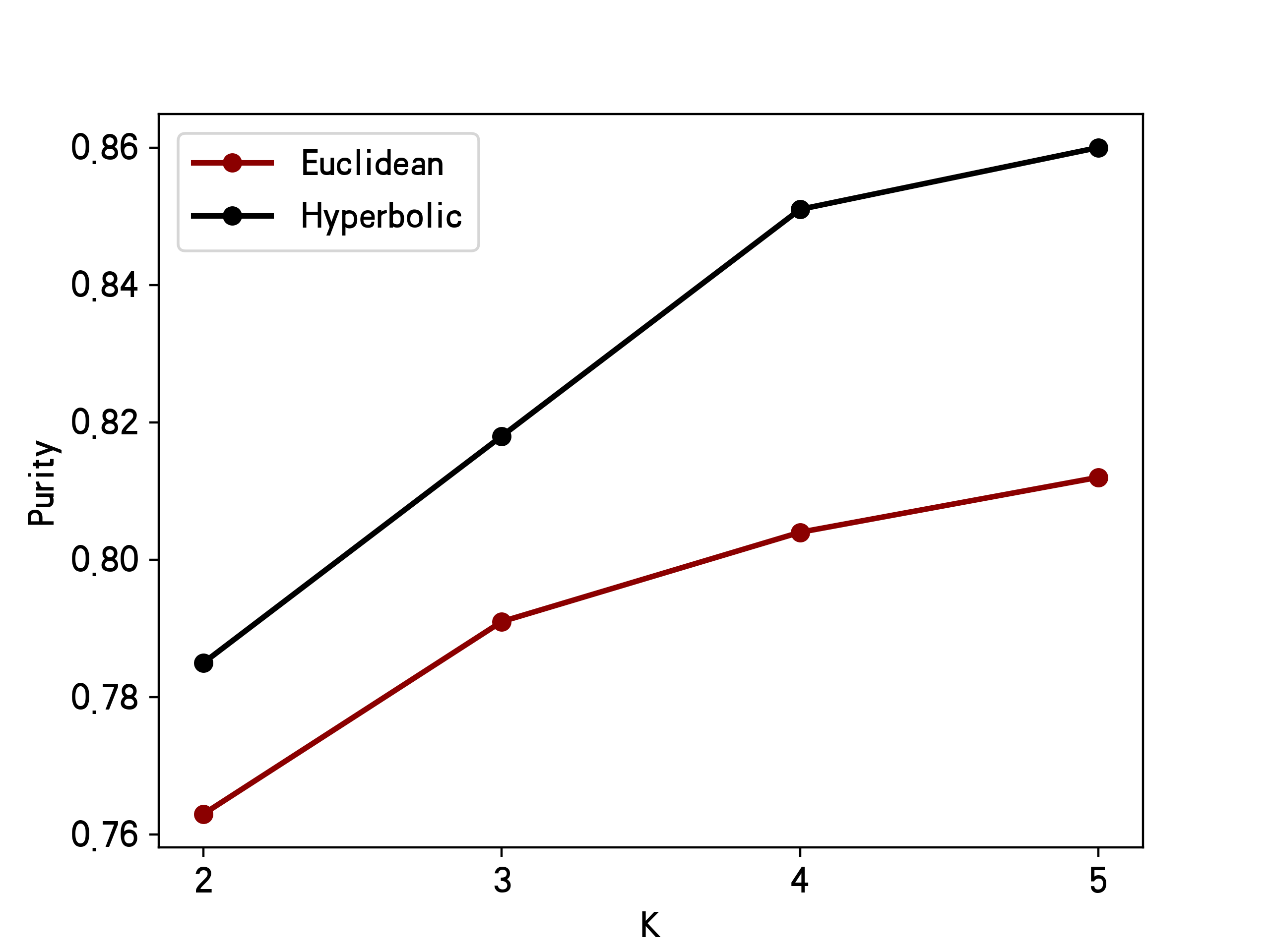}
		\caption{Reconstruction of K} 
        \label{fig:structure_k} 
	\end{minipage}
	\begin{minipage}{0.33\linewidth}
		\includegraphics[width=1.0\linewidth]{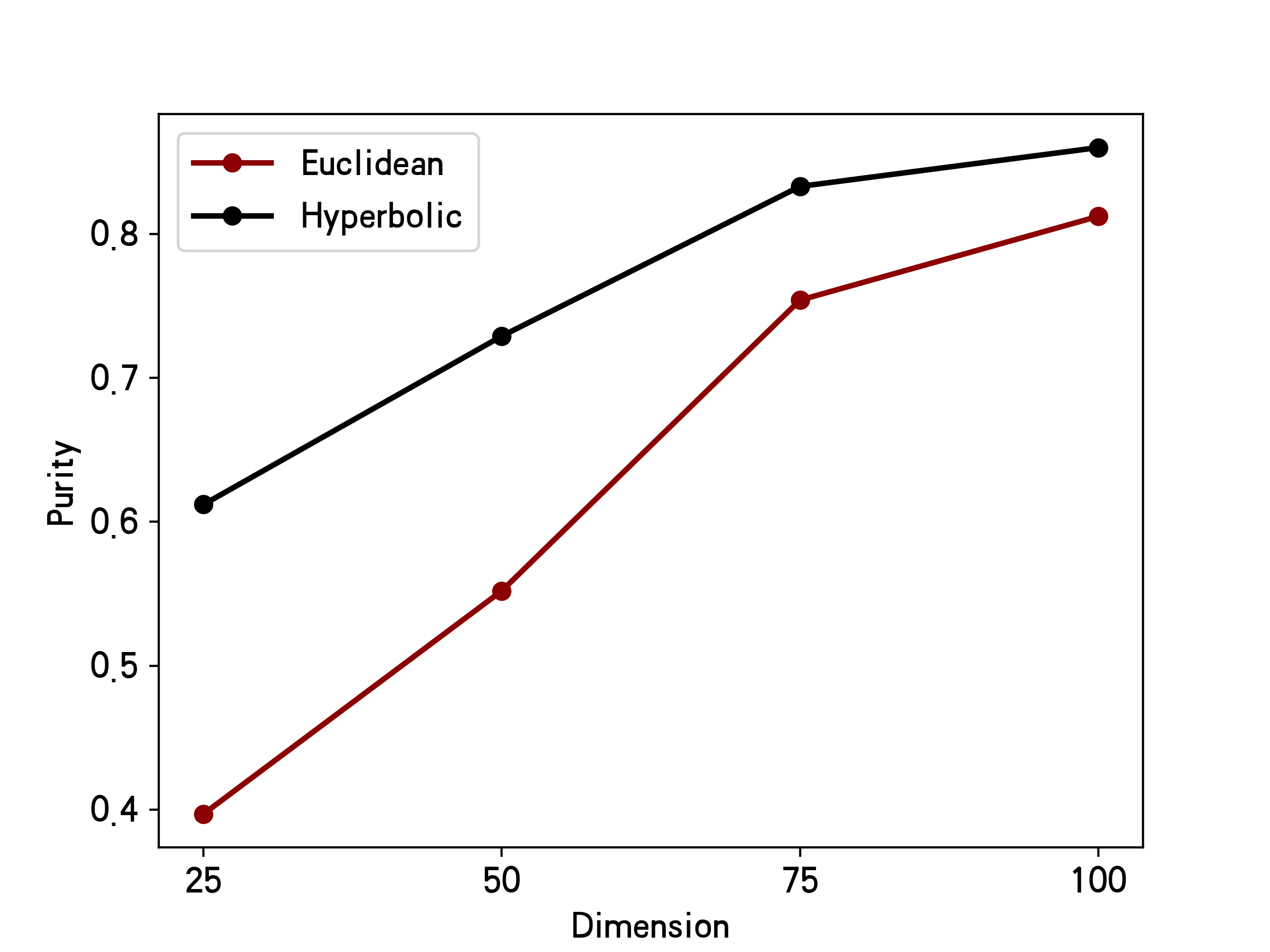}
		\caption{Reconstruction of Dimension} 
        \label{fig:structure_d} 
	\end{minipage}
	\begin{minipage}{0.33\linewidth}
		\includegraphics[width=1.0\linewidth]{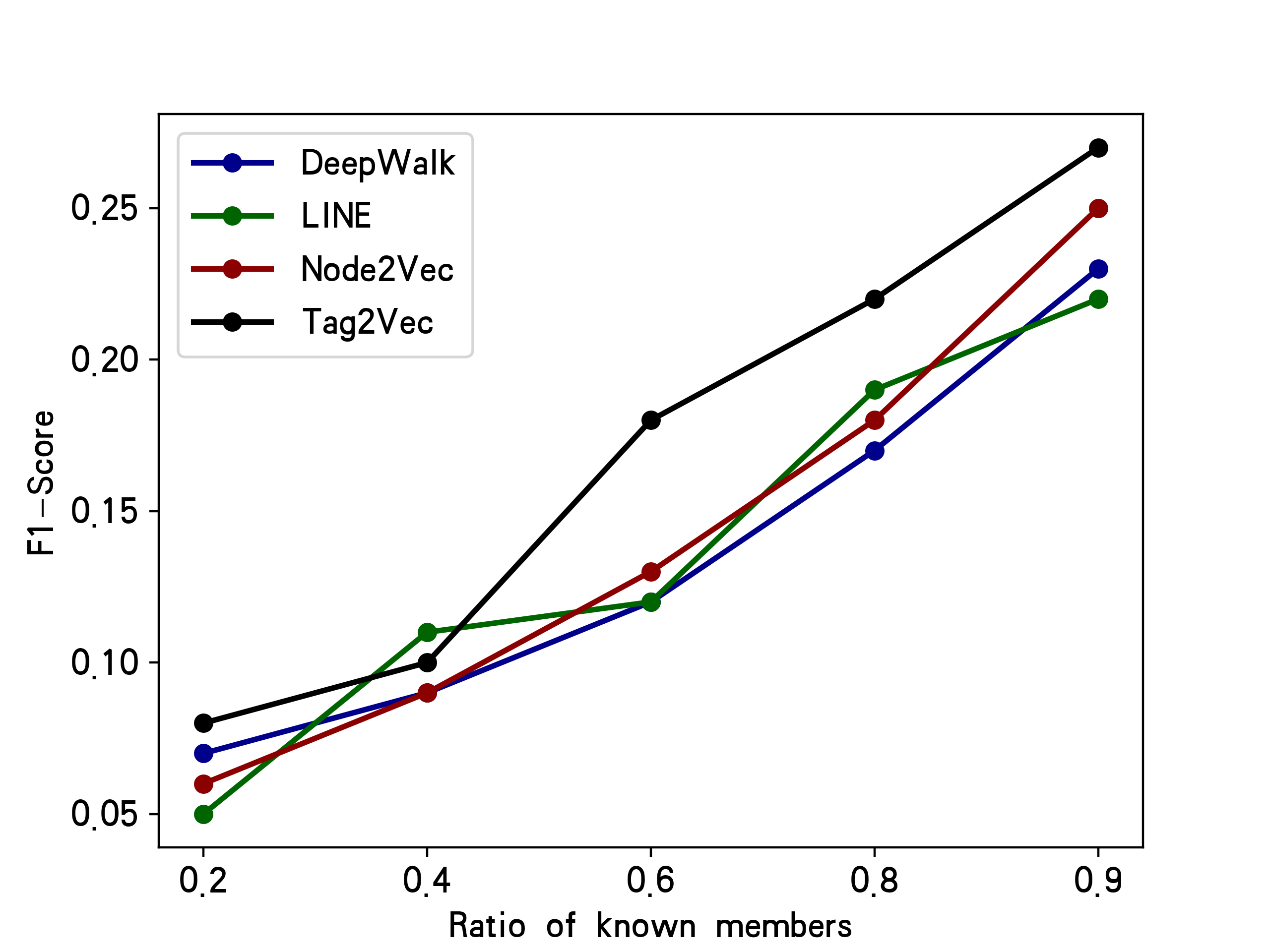}
		\caption{Stability} 
        \label{fig:stability} 
	\end{minipage}
\end{figure*}

\subsection{Results}

\subsubsection{Node Classification}
On NBER patent citation dataset, we concatenate the node and tag vectors together as the expended representations to classify the nodes. The classes and sub-categories are tags of nodes. We use F1 score as the evaluation metric.

Figure \ref{fig:node_classification} shows the results of node classification. It can be seen that origin has the worst result. This is because it encodes tags into one-hot vectors so that all tags have the same status and their distances are ignored. While the rest methods mix the nodes and tags into a hybrid network to learn the relationships between tags. An exceptional case is that LINE not performs well because it ignores the global information. Furthermore, our model outperforms DeepWalk and Node2Vec, as parametrized random walk can better learn the semantic information, and hyperbolic Skip-gram helps to learn the hierarchical information. It demonstrates that the representations learned from our model have stronger ability for classification.

\subsubsection{Similar Community Detection}
We detect the similar communities on the NBER patent citation dataset. A community corresponds to a tag, including classes, categories and sub-categories. We calculate distances between tag vectors. As for a community, the community with closest distance is its similar community. The similar community refers to the one with most common members or most connections. We use AUC as the evaluation metric.

Figure \ref{fig:similar_community_detection} shows the results of similar community detection. For all baselines, we mix the nodes and tags into a hybrid network. The representation vectors learned from our model can predict the similarity relationships between communities and analyze the community features of complex networks.

\subsubsection{Hierarchical Structure Reconstruction}
Comparing with Euclidean spaces, Skip-gram model in hyperbolic spaces has two advantages: reducing the dimensions of vectors and reflecting affiliation relationships better. To evaluate this, we conduct the experiment of hierarchical structure reconstruction on WordNet dataset to compare the performance of representation vectors from Euclidean and hyperbolic spaces. We reconstruct the word tree by clustering the tag representation vectors and evaluate the results with purity $\sum_{i}^K \max_j \{p_{ij}\},$ where $K$ is cluster number, and $p_{ij}$ is the probability that members in cluster $i$ belong to class $j$. Purity measures the correctness of hierarchical structure reconstruction. If we can cluster correctly, it means that we can discover the complex hierarchical structure of ground-truth communities of tags with the help of the behaviors of their members in the network. 

Figure \ref{fig:structure_k} shows the purity of clustering with different order of ancestors $k$ in two spaces. $k$ measures the richness of tags and the complexity of hierarchical information. The larger the $k$ is, the more tags a node has. The result shows that embedding in hyperbolic spaces is superior with the best reconstruction performance. Meanwhile, when there are more tags and more complex structure, the advantage of hyperbolic spaces is greater.

Figure \ref{fig:structure_d} shows the purity of clustering with different dimensions. When the dimensions are lower, embedding in hyperbolic spaces performs significantly better than embedding in Euclidean spaces. It indicates that in hyperbolic spaces, we can learn good representations with low dimensions.

\subsubsection{Tag Stability}
In real networks, tags of nodes often make changes. Besides, part of interaction records may be missing. However, we consider that changes or lacking of few members may not influence the semantic information of tags greatly. To demonstrate the stability of tags learned from our model, we extract part of plain nodes as known members and all tag nodes to construct a network and learn the tag representations. Then we extract another part of plain nodes to construct another network without tags, learn the representation vectors of plain nodes and combine them with tag vectors learned before to predict the node labels.

The results are shown in Figure \ref{fig:stability}. When the known members reach a certain proportion(60\%), F1-score of TagVec is higher than other baselines obviously. This is because in the parameterized random walk, we emphasize the importance of tags when generating context. When some members are missing, we can still learn good tag representation vectors thanks to their stronger stability. 

\section{Conclusion}
In this paper, we analyze semantic and hierarchical information between tags and propose a novel tag representation learning model, Tag2Vec. The experimental results show that the tag representations learned from our model have strong representative ability.

\begin{acks}
This work was supported by the National Natural Science Foundation of China (Grant No. 61876006 and No. 61572041).
\end{acks}

\bibliographystyle{main}
\balance 
\bibliography{main}

\end{document}